\documentclass[prl,twocolumn,floatfix]{revtex4}

\usepackage{graphicx}
\usepackage{amsmath}
\usepackage{amssymb}
\usepackage[colorlinks=true,citecolor=blue,linkcolor=blue]{hyperref}
\usepackage{amsfonts}
\usepackage{color,xcolor}
\usepackage{epstopdf}
\usepackage{braket}
\usepackage{bm}
\usepackage{bbold}

\renewcommand{\vec}[1]{\ensuremath{\boldsymbol{#1}}}

\begin{document}

\title{Strong Valley Zeeman Effect of Dark Excitons in Monolayer Transition Metal Dichalcogenides in a Tilted Magnetic Field}
\date{\today}
\author{M. Van der Donck}
\email{matthias.vanderdonck@uantwerpen.be}
\affiliation{Department of Physics, University of Antwerp, Groenenborgerlaan 171, B-2020 Antwerp, Belgium}
\author{M. Zarenia}
\email{mohammad.zarenia@uantwerpen.be}
\affiliation{Department of Physics, University of Antwerp, Groenenborgerlaan 171, B-2020 Antwerp, Belgium}
\author{F. M. Peeters}
\email{francois.peeters@uantwerpen.be}
\affiliation{Department of Physics, University of Antwerp, Groenenborgerlaan 171, B-2020 Antwerp, Belgium}

\begin{abstract}
The dependence of the excitonic photoluminescence (PL) spectrum of monolayer transition metal dichalcogenides (TMDs) on the tilt angle of an applied magnetic field is studied. Starting from a four-band Hamiltonian we construct a theory which quantitatively reproduces the available experimental PL spectra for perpendicular and in-plane magnetic fields. In the presence of a tilted magnetic field, we demonstrate that the dark exciton PL peaks brighten due to the in-plane component of the magnetic field and split for light with different circular polarization as a consequence of the perpendicular component of the magnetic field. This splitting is more than twice as large as the splitting of the bright exciton peaks in tungsten-based TMDs. We propose an experimental setup that will allow to access the predicted splitting of the dark exciton peaks in the PL spectrum.

\end{abstract}

\maketitle

Single layers of semiconducting transition metal dichalcogenides (TMDs) have been the subject of intensive theoretical \cite{theory1,theory2,theoryexp} and experimental \cite{theoryexp,exp1,exp2} research in recent years. These studies have been motivated by several unique features of TMDs: \emph{i}) the lack of inversion symmetry that leads to the formation of a large direct band gap ($\gtrsim 1.5$ eV) at the two inequivalent valleys located at the $K$ and $K'$ points of the hexagonal Brillouin zone, \emph{ii}) strong spin-orbit interaction which significantly lifts the degeneracy between the spin levels of the conduction and valence bands \cite{lambdac1,lambdac2,lambdac3,lambdac4}, and \emph{iii}) strong excitonic effects at room temperature that originate from the two-dimensional (2D) character of TMDs and the associated reduced dielectric screening of the Coulomb interaction between charge carriers \cite{mak,chernikov,he,sallen,korn,exctheory1,exctheory2,exctheory3}.

\begin{figure}
\centering
\includegraphics[width=8.5cm]{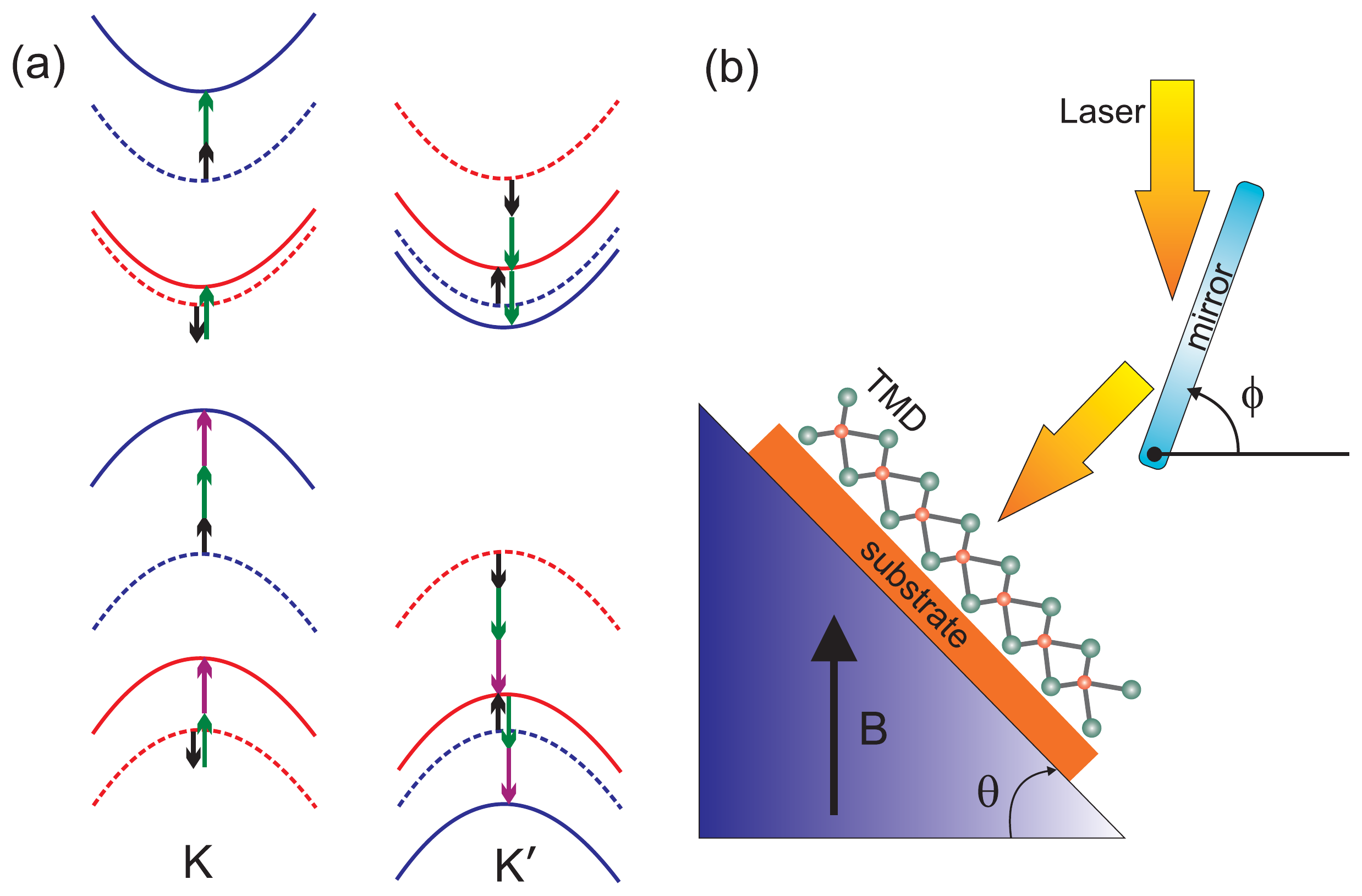}
\caption{(Color online) (a) Schematic representation of the different magnetic shifts of the energy levels of tungsten-based monolayer TMDs in the absence (dashed) and presence (full) of a perpendicular magnetic field. Landau levels are not shown here. Blue and red curves are spin up and spin down bands, respectively. The black, brown, and green arrows indicate the effect of the spin, intracellular orbital, and intercellular orbital magnetic moment, respectively (explained in the text). (b) Schematic representation of an experimental setup for studying the effects of a tilted magnetic field on the optical properties of a TMD monolayer.}
\label{fig:exp}
\end{figure}

The two low-energy valleys in TMDs are degenerate due to time-reversal symmetry. However, the application of an external magnetic field breaks the time-reversal symmetry and as a consequence lifts this degeneracy. This is referred to as the valley Zeeman effect. Using magneto-photoluminescence spectroscopy, the valley Zeeman effect has been experimentally observed in TMD monolayers as different energy shifts induced in the excitonic transitions in the two valleys by a \emph{perpendicular} magnetic field \cite{perpabs1,perpabsn,perpabs2,perpabs3,perpabsnn,perpabs4}. A perpendicular magnetic field, aside from leading to Landau quantization, decreases (increases) the energy gap between the highest valence band state and lowest conduction band state via the intracellular orbital magnetic moment (the magnetic moment of the particles around their atomic site) in the $K$ ($K'$) valley, implying that the exciton transition energy is different in the two valleys. Due to the circular dichroism in TMDs this means that the exciton resonances in the photoluminescence (PL) spectra for left and right circularly polarized light will shift from each other. The different magnetic shifts of the energy levels in the presence of a perpendicular magnetic field are schematically depicted in Fig. \ref{fig:exp}(a) and will be discussed in detail later.

Depending on the relative sign of the spin-orbit couplings in the conduction and valence bands in TMDs, the excitonic ground state can be bright (parallel spin configuration at the lowest conduction and highest valence band for which the optical transition is allowed) or dark (opposite spin configuration and optically forbidden ground state interband transition). An \emph{in-plane} magnetic field, aside from leading to small additional shifts in the energy bands, couples the different spin states and as a result leads to a finite amplitude for previously forbidden interband transitions and as such leads to additional peaks in the PL spectrum. This brightening of dark excitons by an in-plane magnetic field was demonstrated in a recent experiment \cite{parabs}.

In this letter, we investigate the influence of a \emph{tilted} magnetic field on the excitonic PL spectrum in monolayer TMDs. A tilted magnetic field allows to combine the effects of the breaking of the energy degeneracy of the two valleys and the coupling between the different spin states. We predict that the extra peak, which arises due to the in-plane component of the magnetic field, will also split due to the perpendicular magnetic field component and that this splitting is more than twice as large as the splitting of the bright peaks in tungsten-based TMDs. In Fig. \ref{fig:exp}(b) we show a possible experimental setup. The magnetic field is oriented along the $z$-direction and the sample can be tilted over an arbitrary angle $\theta$. The mirror should be tilted over an angle $\phi=(\pi-\theta)/2$ in order to have perpendicular incidence of the laser beam which is pointed along the $z$-direction. Starting from the four-band low-energy dispersion of TMD monolayers we present a semi-analytical approach for calculating the exciton energies and wave functions. The obtained results quantitatively reproduce the experimental PL spectra for perpendicular and in-plane magnetic fields.

We start from the effective low-energy single-electron Hamiltonian \cite{theory1} in the basis $\mathcal{B}^e_{\tau}=\{\ket{\phi^e_{c,\uparrow,\tau}},\ket{\phi^e_{v,\uparrow,\tau}},\ket{\phi^e_{c,\downarrow,\tau}},\ket{\phi^e_{v,\downarrow,\tau}}\}$ spanning the 4D Hilbert space $\mathcal{H}^e_{\tau}$, with $\ket{\phi^e_{c,\uparrow(\downarrow),\tau}}$ and $\ket{\phi^e_{v,\uparrow(\downarrow),\tau}}$ the spin up (down) atomic orbital states at the conduction $(c)$ and valence $(v)$ band edge, respectively, and incorporate an arbitrarily oriented magnetic field:
\begin{widetext}
\begin{equation}
\label{singelham}
\begin{split}
H^q_{\tau}(\vec{\Pi}) = I_2^s\otimes\left(at\vec{\Pi}^{\tau}.\vec{\sigma}+\frac{\Delta}{2}\sigma_z-2\tau\frac{q}{e}\mu_BB_z\frac{I_2^p-\sigma_z}{2}\right)- \left(\frac{q}{e}\mu_B\vec{B}.\vec{s}\right)\otimes I_2^p+\tau s_z\otimes\left(\lambda_c\frac{I_2^p+\sigma_z}{2}+\lambda_v\frac{I_2^p-\sigma_z}{2}\right),
\end{split}
\end{equation}
\end{widetext}
where $\vec{\sigma}$ $(\vec{s})$ is a vector with the pseudospin (spin) Pauli matrices $\sigma_i$ $(s_i)$ ($i=x,y,z$) as its components, $I_2^p$ $(I_2^s)$ is the two by two pseudospin (spin) identity matrix, $a$ the lattice constant, $t$ the hopping parameter, $\tau=\pm1$ the valley index, $\Delta$ the band gap, $\lambda_{c(v)}$ the spin-orbit coupling strength leading to a spin splitting of $2\lambda_{c(v)}$ at the conduction (valence) band edge, $q$ the charge of the electron, $e$ the elementary charge, $\mu_B$ the Bohr magneton, and $\vec{\Pi}^{\tau}=(\tau\Pi_x,\Pi_y,0)^T$ with $\Pi_i=k_i-qA_i/\hbar$ where $\vec{A}$ is the vector potential giving rise to the magnetic field $\vec{B}=\vec{\nabla}\times\vec{A}$. Here, we choose to work in the gauge $\vec{A}=(-B_zy/2,B_zx/2-B_xz,0)^T$ with $B_x=B\sin\theta$ and $B_z=B\cos\theta$ where $B$ is the magnetic field strength. However, since we are considering a 2D system we can take $z=0$. The first part of the above Hamiltonian is the gapped Dirac Hamiltonian plus the contribution of the intracellular orbital magnetic moment (conduction and valence band states in monolayer TMDs have $m_z=0$ and $m_z=2\tau$, respectively). The second and third part are the contribution of the spin magnetic moment and the spin-orbit coupling, respectively.

The in-plane components of the magnetic field prevent the above Hamiltonian from being diagonal in spin space and reducing it to two 2D Hamiltonians. Therefore, in order to have a 4D exciton Hamiltonian instead of a 16D one, we will consider the in-plane components of the magnetic field within first order perturbation theory, which leads to the effective 2D Hamiltonian in the basis $\mathcal{B}^e_{s,\tau}=\{\ket{\phi^e_{c,s,\tau}},\ket{\phi^e_{v,s,\tau}}\}$:
\begin{equation}
\label{singelhameff}
\begin{split}
H^q_{s,\tau}(\vec{\Pi}) = &at(\tau \Pi_x\sigma_x+\Pi_y\sigma_y)+\frac{\Delta}{2}\sigma_z+\tilde{\lambda}_cs\tau\frac{I_2+\sigma_z}{2} \\
&+\left(\tilde{\lambda}_vs\tau-2\tau\frac{q}{e}\mu_BB_z\right)\frac{I_2-\sigma_z}{2}-s\frac{q}{e}\mu_BB_zI_2,
\end{split}
\end{equation}
with $s=\pm1$ the spin index and with $\tilde{\lambda}_{c(v)}=\lambda_{c(v)}+\mu_B^2B_x^2/(2\lambda_{c(v)})$. This is a good approximation as long as $\mu_BB_x$ is small compared to $\lambda_c$.

Since a hole with wave vector $\vec{k}$, spin $s$, and valley index $\tau$ can be described as the absence of an electron with opposite wave vector, spin, and valley index, the single-hole Hamiltonian can immediately be obtained from the single-electron Hamiltonian and is given by $-H^{-q}_{-s,-\tau}(-\vec{\Pi})$. The eigenstates of this Hamiltonian span the 2D Hilbert space $\mathcal{H}^h_{s,\tau}$. The total exciton Hamiltonian acts on the product Hilbert space spanned by the tensor products of the single-particle states at the band edges, $\mathcal{B}_{\alpha}=\mathcal{B}^e_{s^e,\tau^e}\otimes\mathcal{B}^h_{s^h,\tau^h}$, and is given by
\begin{equation}
\label{excham}
\begin{split}
H^{\mathrm{exc}}_{\alpha}(\vec{\Pi}^e,\vec{\Pi}^h,r_{eh}) = &H^{q^e}_{s^e,\tau^e}(\vec{\Pi}^e)\otimes I_2 \\
&-I_2\otimes H^{-q^h}_{-s^h,-\tau^h}(-\vec{\Pi}^h)-V(r_{eh})I_{4},
\end{split}
\end{equation}
where $\alpha$ is a shorthand notation for $s^e,\tau^e,s^h,\tau^h$, with $q^h=-q^e=e$, and where the electron-hole interaction potential is given by \cite{screening1,screening2,screening3}
\begin{equation}
\label{inter}
V(r_{ij}) = \frac{e^2}{4\pi\kappa\varepsilon_0}\frac{\pi}{2r_0}\left[H_0\left(\frac{r_{ij}}{r_0}\right)-Y_0\left(\frac{r_{ij}}{r_0}\right)\right],
\end{equation}
with $r_{ij}=|\vec{r}_i-\vec{r}_j|$, where $Y_0$ and $H_0$ are the Bessel function of the second kind and the Struve function, respectively, with $\kappa=(\varepsilon_1+\varepsilon_2)/2$ where $\varepsilon_{1(2)}$ is the dielectric constant of the environment above (below) the TMD monolayer, and with $r_0=2\pi\chi_{2\text{D}}/\kappa$ the screening length where $\chi_{2\text{D}}$ is the 2D polarizability of the TMD layer. In this letter we consider TMDs on a SiO$_2$ substrate with dielectric constant $\varepsilon_2=3.8$ and with vacuum on top, i.e. $\varepsilon_1=1$. The eigenvalue problem for the exciton Hamiltonian (\ref{excham}) is now reduced to a set of four coupled equations. The details on how to solve this eigenvalue problem are given in the Supplemental Material [\onlinecite{supplemental}]. When both the excitonic energy spectrum and the wave functions are obtained we can also calculate the PL spectrum using the formula \cite{absorbform}
\begin{equation}
\label{absorbance}
\alpha_{\pm}(\omega) \propto \text{Im}\left(\sum_{s^e,\tau^e,s^h,n}e^{-\frac{E_{\alpha,n}}{k_BT}}\frac{|\mathcal{P}_{\pm}^{s^e\tau^e}|^2|\phi_{c,v,\alpha,n}^{e,h}(0,0)|^2}{\omega\left(E_{\alpha,n}-\hbar\omega-i\gamma\right)}\right),
\end{equation}
with $\mathcal{P}_{\pm}^{s^e\tau^e}$ the transition amplitude between the single-particle states (for which an expression is derived in the Supplemental Material [\onlinecite{supplemental}]), $E_{\alpha,n}$ the exciton energy of the $n^\text{th}$ state with indices $\alpha$, $\phi_{c,v,\alpha,n}^{e,h}$ the corresponding dominant component of the exciton wave function, $\hbar\omega$ the photon energy, $\gamma$ the broadening of the peaks, and where the hole valley index is fixed for optical transitions at $\tau^h=-\tau^e$.
\begin{figure}
\centering
\includegraphics[width=8.5cm]{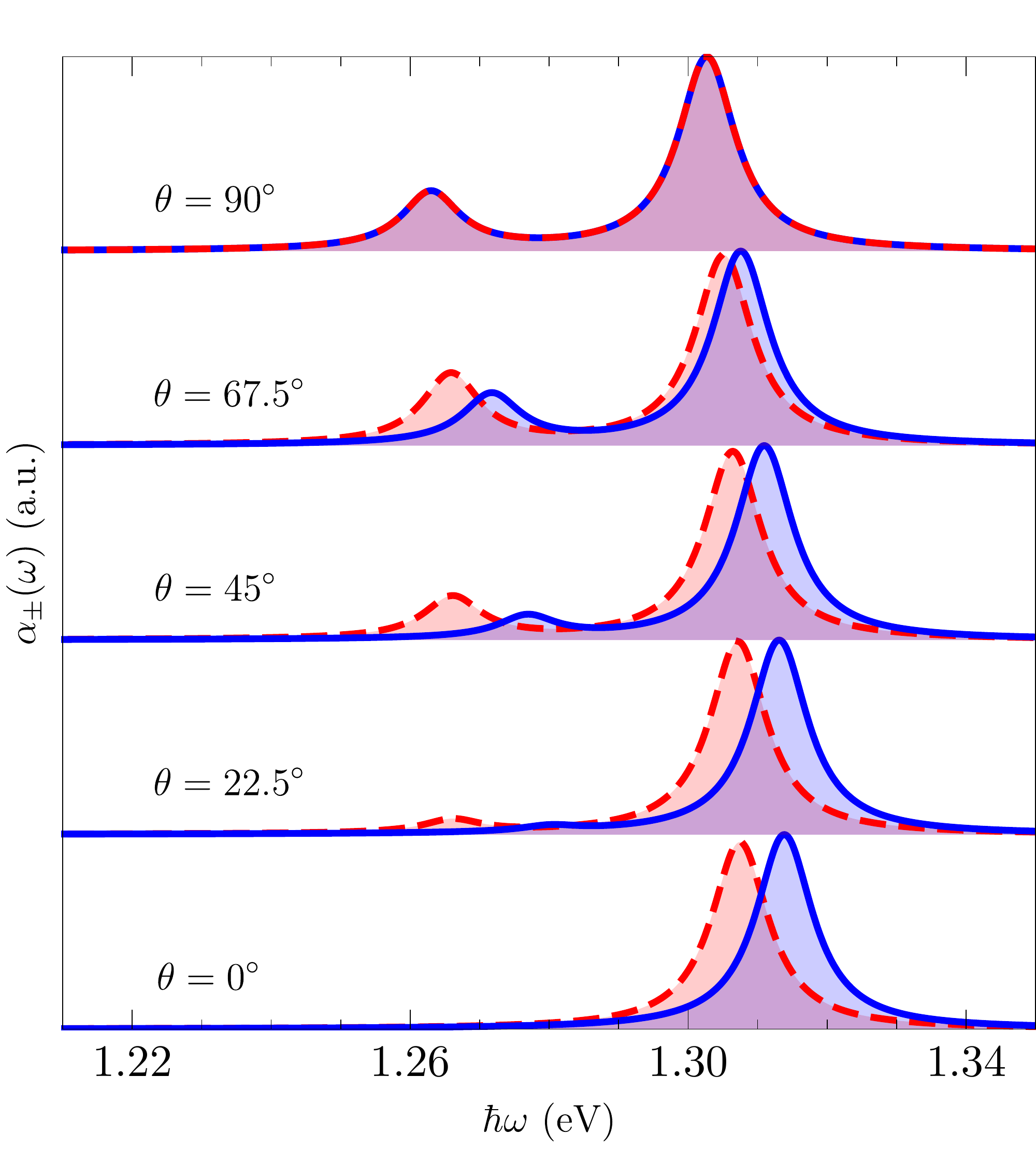}
\caption{(Color online) Excitonic PL spectra of WS$_2$ on a SiO$_2$ substrate for $\sigma_-$ (blue, solid) and $\sigma_+$ (red, dashed) circularly polarized light for different tilt angles of the sample in the presence of a magnetic field of 30 T. We used a broadening of $\gamma=5$ meV.}
\label{fig:pl}
\end{figure}
\begin{figure}
\centering
\includegraphics[width=7cm]{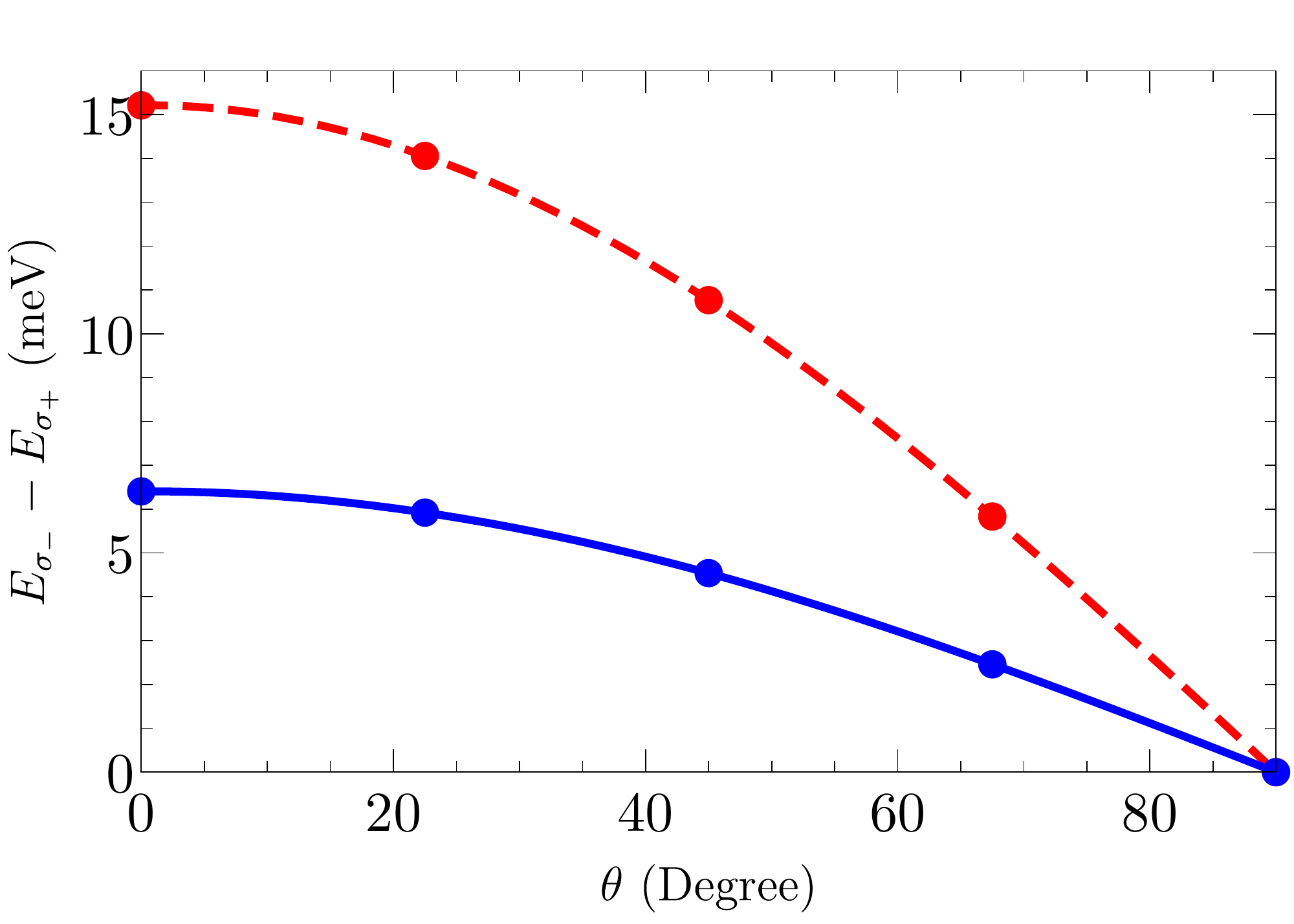}
\caption{(Color online) Splitting of the bright (blue, solid) and dark (red, dashed) excitonic peaks in the PL spectrum of WS$_2$ on a SiO$_2$ substrate in the presence of a magnetic field of 30 T as a function of the tilt angle of the sample.}
\label{fig:splitting}
\end{figure}

As mentioned in the introduction, the intracellular orbital magnetic moment leads to different energy gaps in the two valleys. More specifically it decreases (increases) the energy gap in the $K$ ($K'$) valley by an amount of $2\mu_BB_z$. For states with the same spin and valley index, we can see from Fig. \ref{fig:exp}(a) that the intercellular orbital magnetic moment (the intrinsic magnetic moment of the individual Bloch particles for which an expression is given in the Supplemental Material [\onlinecite{supplemental}]) and spin magnetic moment do not influence the energy gap. As a result, the bright exciton peaks in the PL spectrum split by an amount of $4\mu_BB_z$ between the two circular polarizations of the laser. When $\lambda_c$ and $\lambda_v$ have the same sign the ground state of the $A$ exciton, i.e. an exciton in which the hole stems from the highest valence band, is dark. Theoretical studies predict that this is the case for TMDs consisting of tungsten, while it is not the case for TMDs consisting of molybdenum \cite{lambdac1,lambdac2,lambdac3,lambdac4}. Here we assume $\lambda_v>0$ and as a consequence $\lambda_c>0$ for tungsten based TMDs and $\lambda_c<0$ for molybdenum based TMDs. However, in the presence of an in-plane magnetic field these dark excitons become brightened. Fig. \ref{fig:exp}(a) shows that the energy gap between the highest valence band and the conduction band with opposite spin increases (decreases) with $4\mu_BB_z$ in the $K$ ($K'$) valley due to the spin and intracellular orbital magnetic moments. The intercellular orbital magnetic moment will further add to this difference in size of the energy gap since it has a different magnitude for different spin states. Therefore, in a tilted magnetic field the peaks in the PL spectrum due to these brightened dark excitonic states will also split between the two circular polarizations of the laser and this splitting is expected to be more than twice as large as the splitting between the bright exciton peaks. For materials with $\lambda_c>0$ the dark exciton energy is lower than the bright exciton energy and therefore these resonances can be detected in the PL spectrum as their intensity is further thermally increased by a factor $\text{exp}[\Delta E_{bd}/(k_BT)]$, with $\Delta E_{bd}$ the difference between the bright and the dark exciton energy.

This is illustrated in Fig. \ref{fig:pl} where we show the excitonic PL spectrum of WS$_2$ for different tilt angles. The results clearly show the above predicted effects, with the splitting of the dark exciton peaks more than twice as large as compared to the splitting of the bright exciton peaks, which should be detectable experimentally. However, although the splitting of the dark exciton peak increases as the tilt angle decreases, the intensity of the dark exciton peaks decreases as well, making them more difficult to observe. Therefore, this effect can be best measured at intermediate tilt angles. The splitting of the bright and dark excitonic peaks in the PL spectrum is shown in Fig. \ref{fig:splitting} as a function of the tilt angle of the sample. Notice that the splitting of both excitonic peaks increases with decreasing angle and that the splitting of the dark excitonic peak is more than twice as large as compared to the splitting of the bright peak.
\begin{figure}
\centering
\includegraphics[width=8.5cm]{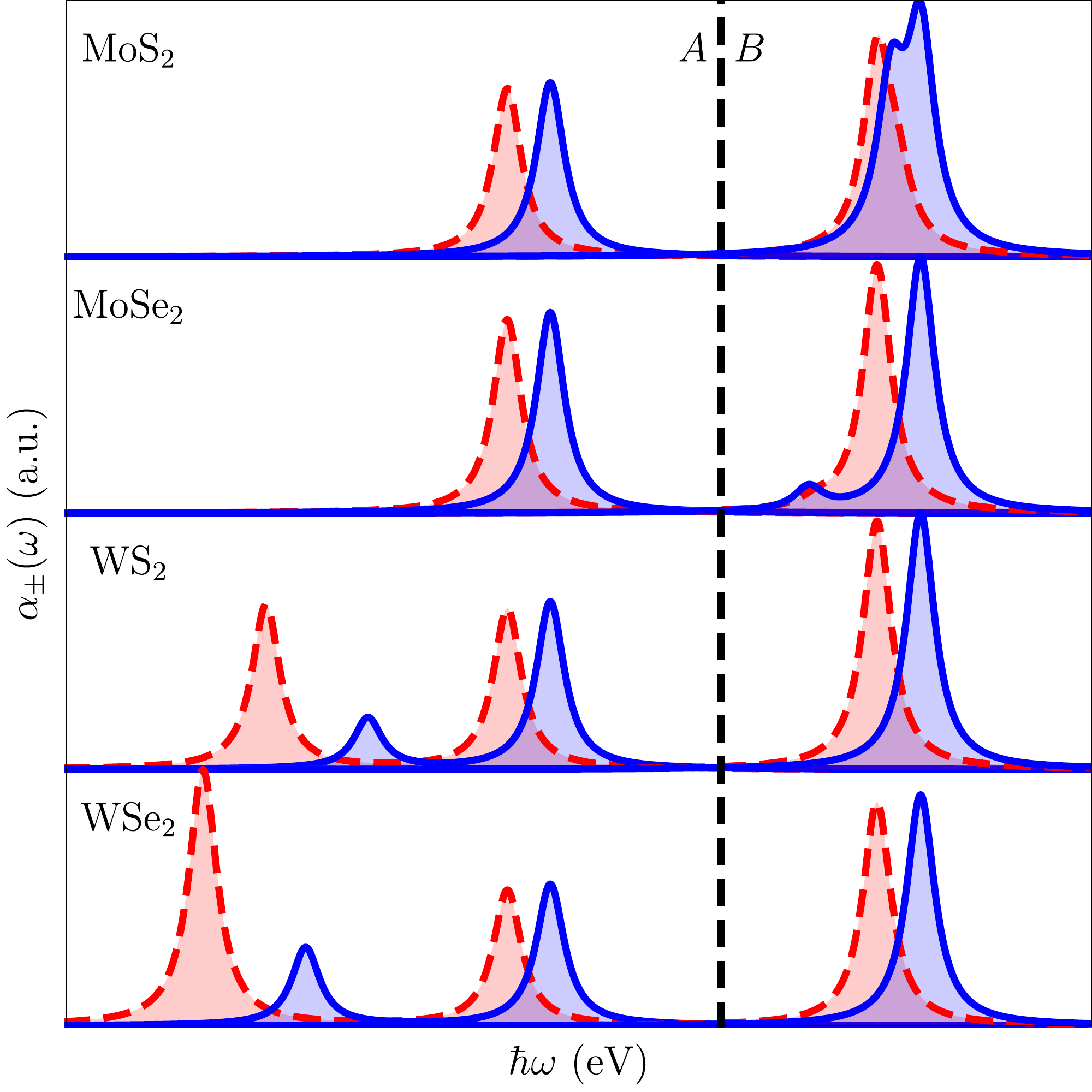}
\caption{(Color online) Schematic excitonic PL spectra for $\sigma_-$ (blue, solid) and $\sigma_+$ (red, dashed) circularly polarized light for different TMD monolayers on a SiO$_2$ substrate with tilt angle $\theta=45^{\circ}$ in the presence of a magnetic field of 50 T. We used a broadening of $\gamma=3$ meV.}
\label{fig:plmat}
\end{figure}

For materials with $\lambda_c<0$ the dark exciton energy is higher than that of the bright exciton and as such these states are, in addition to their already lower intensity, further thermally suppressed by a factor $\text{exp}[\Delta E_{bd}/(k_BT)]$ and are therefore not detected experimentally. This can be seen in Fig. \ref{fig:plmat}, where no brightened dark $A$ exciton peaks are seen in the PL spectrum of MoS$_2$ and MoSe$_2$. For $B$ excitons, i.e. excitons in which the hole stems from the lowest valence band, the situation is reversed: in materials with $\lambda_c<0$ the dark exciton has a lower energy than the bright exciton and can be detected whereas in materials with $\lambda_c>0$ the dark exciton has a higher energy than the bright exciton and is thermally suppressed. However, in this case the spin and intracellular orbital magnetic moments cancel each other and the only change in the energy gap comes from the intercellular orbital magnetic moment. Therefore, the splitting of the dark $B$ exciton peaks in the PL spectrum will be smaller than that of the bright excitons and thus more difficult to detect. This can be seen in the figure, where the brightened dark $B$ exciton peak of MoSe$_2$ and that of MoS$_2$ are difficult to observe. Although, for the latter, the treatment of the in-plane component of the magnetic field within first order perturbation theory might have smaller accuracy due to the very small $\lambda_c$.

The material constants for four different TMDs used in this work are listed in Table \ref{table:mattable}. Changing these values would only lead to shifts in the PL spectra. The only parameter which is of qualitative importance is (the sign of) $\lambda_c$. Furthermore, we only consider excitons in the $1s$-state in the results presented here.

\begin{figure}
\centering
\includegraphics[width=8.5cm]{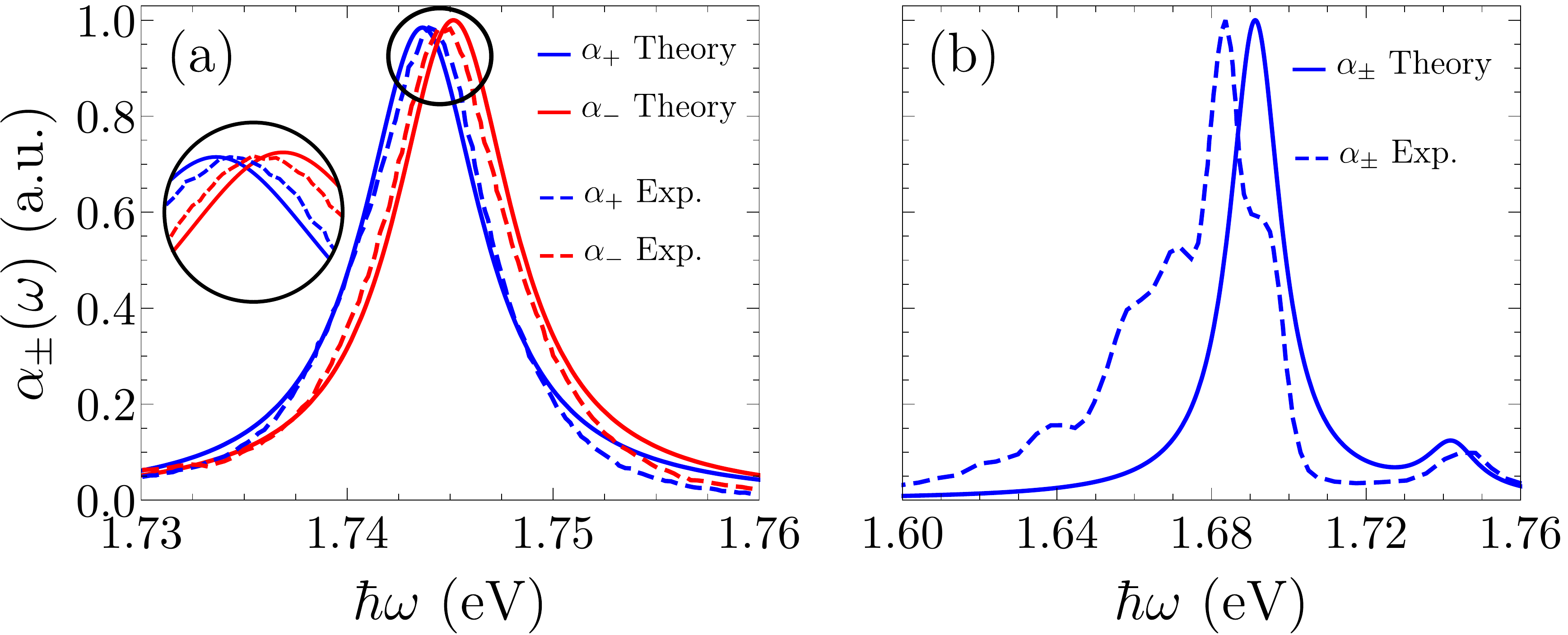}
\caption{(Color online) Excitonic PL spectra for $\sigma_+$ and $\sigma_-$ circularly polarized light for WSe$_2$ on a SiO$_2$ substrate for $B=7$ T, $\theta=0^{\circ}$, $\gamma=3.5$ meV (a) and $B=14$ T, $\theta=90^{\circ}$, $\gamma=8$ meV (b). The results of our model are shifted to match the $A$ exciton energy of the experimental results and the maxima are rescaled to facilitate comparison.}
\label{fig:expcomp}
\end{figure}

In Figs. \ref{fig:expcomp}(a) and (b) we compare our results with experimental results for the case of a perpendicular \cite{perpabs2} and a parallel \cite{parabs} magnetic field, respectively. For a perpendicular magnetic field we find a slightly larger splitting of the excitonic peak, which is possibly due to the fact that the magnetic quantum numbers in the conduction and valence bands of monolayer TMDs deviate somewhat from the values $m_z=0$ and $m_z=2\tau$ due to mixing of the $d$ orbitals that make up the single-particle states at the band edges with $p$ orbitals \cite{perpabs4}. In the case of a parallel magnetic field there are additional features in the experimental PL spectrum which have been attributed to localized or defect-related excitons, as well as to trions \cite{defect1,defect2}. The defect-related exciton could be studied by adding a Coulomb-like impurity to the diagonal elements of the single-particle Hamiltonian \eqref{singelham}. Trions and biexcitons pose a considerably bigger challenge. Constructing a trion or biexciton Hamiltonian can be done in a similar fashion as described here for the exciton. However, solving the corresponding eigenvalue problem would be computationally impossible since in these cases the angular correlations can not be neglected \cite{compare} and this would require solving a 6D (trion) and 8D (biexciton) differential equation.

In summary, we have constructed a theory which allows to calculate the effect of an arbitrarily oriented magnetic field on excitons in monolayer transition metal dichalcogenides. We found that for tungsten based TMDs the dark $A$ exciton peak in the PL spectrum, which is brightened due to the in-plane component of the magnetic field, splits between left and right circularly polarized light due to the perpendicular component of the magnetic field and that this splitting is more than twice as large as compared to the splitting of the bright exciton peak, which should be observable experimentally.
\begin{table}
\centering
\caption{Lattice constants \cite{theory1}, hopping parameters \cite{theory1}, band gaps \cite{theory1}, spin splittings of the conduction \cite{lambdac2} and valence \cite{lambdav} band, and screening lengths \cite{screeninglength} for different TMD materials suspended in vacuum.}
\begin{tabular}{c c c c c c c}
\hline
\hline
 & $a$ (\AA) & $t$ (eV) & $\Delta$ (eV) & $2\lambda_c$ (meV)& $2\lambda_v$ (meV) & $r_0$ (\AA) \\
\hline
\hline
Mo$\text{S}_2$ & 3.193 & 1.10 & 1.66 & -3 & 150 & 41.47 \\
\hline
MoS$\text{e}_2$ & 3.313 & 0.94 & 1.47 & -21 & 180 & 51.71 \\
\hline
W$\text{S}_2$ & 3.197 & 1.37 & 1.79 & 27 & 430 & 37.89 \\
\hline
WS$\text{e}_2$ & 3.310 & 1.19 & 1.60 & 38 & 460 & 45.11 \\
\hline
\hline
\end{tabular}
\label{table:mattable}
\end{table}

\begin{acknowledgments}
This work was supported by the Research Foundation of Flanders (FWO-Vl) through an aspirant research grant for MVDD and by the Methusalem foundation of the Flemish Government.
\end{acknowledgments}


\begin{thebibliography}{10}
%
\bibitem{theory1}
D. Xiao,  G.-B. Liu, W. Feng, X. Xu, and W. Yao, Phys. Rev. Lett. {\bf 108}, 196802 (2012).

\bibitem{theory2}
M. Danovich, V. Z\'olyomi and V. I. Fal'ko, Sci. Rep. \textbf{7}, 45998 (2017).

\bibitem{theoryexp}
E. Courtade, M. Semina, M. Manca, M. M. Glazov, C. Robert, F. Cadiz, G. Wang, T. Taniguchi, K. Watanabe, M. Pierre, W. Escoffier, E. L. Ivchenko, P. Renucci, X. Marie, T. Amand, and B. Urbaszek, Phys. Rev. B \textbf{96}, 085302 (2017).

\bibitem{exp1}
Q. H. Wang, K. Kalantar-Zadeh, A. Kis, J. N. Coleman, and M. S. Strano, Nat. Nanotechnol. \textbf{7}, 699 (2012).

\bibitem{exp2}
G. Moody and S. T. Cundiff, Adv. Phys.: X \textbf{2}, 641 (2017).

\bibitem{lambdac1}
G.-B. Liu, W.-Y. Shan, Y. Yao, W. Yao, and D. Xiao, Phys. Rev. B \textbf{88}, 085433 (2013).

\bibitem{lambdac2}
K. Ko\'smider, J. W. Gonz\'alez, and J. Fern\'andez-Rossier, Phys. Rev. B \textbf{88}, 245436 (2013).

\bibitem{lambdac3}
A. Korm\'anyos, G. Burkard, M. Gmitra, J. Fabian, V. Z\'olyomi, N. D. Drummond, and V. Fal'ko, 2D Mater. \textbf{2}, 022001 (2015).

\bibitem{lambdac4}
J. P. Echeverry, B. Urbaszek, T. Amand, X. Marie, and I. C. Gerber, Phys. Rev. B \textbf{93}, 121107(R) (2016).

\bibitem{mak}
K. F. Mak, K. He, C. Lee, G. H. Lee, J. Hone, T. F. Heinz, and J. Shan, Nat. Mater. {\bf 12}, 207 (2013).

\bibitem{chernikov}
A. Chernikov, T. C. Berkelbach, H. M. Hill, A. Rigosi, Y. Li, O. B. Aslan, D. R. Reichman, M. S. Hybertsen, and T. F. Heinz, Phys. Rev. Lett. {\bf 113}, 076802 (2014).

\bibitem{he}
K. He, N. Kumar, L. Zhao, Z. Wang, K. F. Mak, H. Zhao, and J. Shan, Phys. Rev. Lett. {\bf 113}, 026803 (2014)

\bibitem{sallen}
G. Sallen, L. Bouet, X. Marie, G. Wang, C. R. Zhu, W. P. Han, Y. Lu, P. H. Tan, T. Amand, B. L. Liu, and B. Urbaszek, Phys. Rev. B {\bf 86}, 081301 (2012).

\bibitem{korn}
T. Korn, S. Heydrich, M. Hirmer, J. Schmutzler, and C. Sch\"uller, Appl. Phys. Lett. {\bf 99}, 102109 (2011).

\bibitem{exctheory1}
G. Bergh\"{a}user and E. Malic, Phys. Rev. B {\bf 89}, 125309 (2014).

\bibitem{exctheory2}
S. Konabe and S. Okada, Phys. Rev. B {\bf 90}, 155304 (2014).

\bibitem{exctheory3}
Y. Ferreiros and A. Cortijo, Phys. Rev. B {\bf 90}, 195426 (2014).

\bibitem{perpabs1}
Y. Li, J. Ludwig, T. Low, A. Chernikov, X. Cui, G. Arefe, Y. D. Kim, A. M. van der Zande, A. Rigosi, H. M. Hill, S. H. Kim, J. Hone, Z. Li, D. Smirnov, and T. F. Heinz, Phys. Rev. Lett. \textbf{113}, 266804 (2014).

\bibitem{perpabsn}
D. MacNeill, C. Heikes, K. F. Mak, Z. Anderson, A. Korm\'anyos, V. Z\'olyomi, J. Park, and D. C. Ralph, Phys. Rev. Lett. \textbf{114}, 037401 (2015).

\bibitem{perpabs2}
G. Aivazian, Z. Gong, A. M. Jones, R.-L. Chu, J. Yan, D. G. Mandrus, C. Zhang, D. Cobden, W. Yao, and X. Xu, Nat. Phys. \textbf{11}, 148 (2015).

\bibitem{perpabs3}
A. Srivastava, M. Sidler, A. V. Allain, D. S. Lembke, A. Kis, and A. Imamo$\breve{\text{g}}$lu, Nat. Phys. \textbf{11}, 141 (2015).

\bibitem{perpabsnn}
G. Wang, L. Bouet, M. M. Glazov, T. Amand, E. L. Ivchenko, E. Palleau, X. Marie, and B. Urbaszek, 2D Mater. \textbf{2}, 034002 (2015).

\bibitem{perpabs4}
G. Plechinger, P. Nagler, A. Arora, A. G. del \'Aguilla, M. V. Ballottin, T. Frank, P. Steinleitner, M. Gmitra, J. Fabian, P. C. M. Christianen, R. Bratschitsch, C. Sch\"uller, and T. Korn, Nano Lett. \textbf{16}, 7899 (2016).

\bibitem{parabs}
M. R. Molas, C. Faugeras, A. O. Slobodeniuk, K. Nogajewski, M. Bartos, D. M. Basko, and M. Potemski, 2D Mater. \textbf{4}, 021003 (2017).

\bibitem{screening1}
A. V. Chaplik and M. V. Entin, Zh. Eksp. Teor. Fiz. \textbf{61}, 2496 (1971).

\bibitem{screening2}
L. V. Keldysh, JETP Lett. \textbf{29}, 658 (1979).

\bibitem{screening3}
P. Cudazzo, I. V. Tokatly, and A. Rubio, Phys. Rev. B \textbf{84}, 085406 (2011).

\bibitem{supplemental}
See Supplemental Material at http://link.aps.org/
supplemental/10.1103/PhysRevB.97.081109 for further technical details. This includes Refs. [\onlinecite{decouple1,decouple2,decouple3,decouple4,compare,blochmagn,theory1}].

\bibitem{absorbform}
M. Kira and S. W. Koch, Progress in Quantum Electronics \textbf{30}, 155 (2006).

\bibitem{lambdav}
Z. Y. Zhu, Y. C. Cheng, and U. Schwingenschl\"ogl, Phys. Rev. B \textbf{84}, 153402 (2011).

\bibitem{screeninglength}
T. C. Berkelbach, M. S. Hybertsen, and D. R. Reichman, Phys. Rev. B {\bf 88}, 045318 (2013).

\bibitem{defect1}
G. Wang, L. Bouet, D. Lagarde, M. Vidal, A. Balocchi, T. Amand, X. Marie, and B. Urbaszek, Phys. Rev. B \textbf{90}, 075413 (2014).

\bibitem{defect2}
G. Plechinger, P. Nagler, J. Kraus, N. Paradiso, C. Strunk, C. Sch\"uller, and T. Korn, Phys. Status Solidi RRL \textbf{9}, 457 (2015).

\bibitem{compare}
M. Van der Donck, M. Zarenia, F. M. Peeters, Phys. Rev. B \textbf{96}, 035131 (2017).

\bibitem{decouple1}
J. Sabio, F. Sols, and F. Guinea, Phys. Rev. B \textbf{81}, 045428 (2010).

\bibitem{decouple2}
O. L. Berman, R. Y. Kezerashvili, and K. Ziegler, Phys. Rev. B \textbf{85}, 035418 (2012).

\bibitem{decouple3}
O. L. Berman and R. Y. Kezerashvili, Phys. Rev. B \textbf{93}, 245410 (2016).

\bibitem{decouple4}
O. L. Berman, R. Y. Kezerashvili, and K. Ziegler, Phys. Rev. A \textbf{87}, 042513 (2013).

\bibitem{blochmagn}
D. Xiao, M.-C. Chang, and Q. Niu, Rev. Mod. Phys. \textbf{82}, 1959 (2010).

\end{thebibliography}
\end{document}